\documentclass{elsart}

\usepackage{harvard}

\usepackage{graphicx}

\usepackage{amssymb}

\begin{document}

{\Large \bf On the Measurement of the Lense-Thirring effect
using the nodes of the LAGEOS satellites, in reply to ``On the reliability of the so-far performed
tests for measuring the Lense–-Thirring effect with the LAGEOS
satellites'' by L. Iorio}
\footnote{doi of the paper by L. Iorio: 10.1016/j.newast.2005.01.001}\\

 \vspace{28pt}

\begin{centering}
I. Ciufolini$^a$ and E. Pavlis$^b$\\

\end{centering}
\vspace{.25in}

$^a$Dipartimento di Ingegneria dell'Innovazione, Universit\`{a} di
Lecce, Via Monteroni, 73100 Lecce, Italy\\

$^b$Joint Center for Earth Systems Technology (JCET/UMBC),
University of Maryland, Baltimore County, 1000 Hilltop Circle, Baltimore, Maryland, USA 21250\\

{\Large \bf Abstract}

In this paper, we provide a detailed description of our recent
analysis and determination of the frame-dragging effect obtained
using the nodes of the satellites LAGEOS and LAGEOS 2, in reply to
the paper ``On the reliability of the so-far performed tests for
measuring the Lense–Thirring effect with the LAGEOS satellites'' by
L. Iorio (doi: 10.1016/j.newast.2005.01.001). First, we discuss the
impact of the $\dot J_{2n}$ uncertainties on our measurement and we
show that the corresponding error is of the order of 1 $\%$ of
frame-dragging only. We report the result of the orbital simulations
and analyses obtained with and without $\dot J_4$ and a with $\dot
J_4$ equal to its EIGEN-GRACE02S value plus 12 times its published
error, i.e., a $\dot J_4$ equal to about 611 $\%$ of the value
adopted in EIGEN-GRACE02S, that is $\dot J_4 = 6.11 \times
(-1.41\cdot10^{-11}) \simeq -8.61\cdot10^{-11}$. In all these three
cases, by also fitting the final combined residuals with a
quadratic, we obtain the same value of the measured Lense-Thirring
effect. This value differs by only 1 $\%$ with respect to our recent
measurement of the Lense-Thirring effect. Therefore, the error due
to the uncertainties in $\dot J_{2n}$ in our measurement of the
gravitomagnetic effect can at most reach 1 $\%$, in complete
agreement with our previously published error budget. Our total
error budget in the measurement of frame-dragging is about $5 \%$ of
the Lense-Thirring effect, alternatively even by simply considering
the published errors in the $\dot J_{2n}$ and their recent
determinations we get a total error budget of the order of 10 $\%$,
in complete agreement with our previously published error budget.
Furthermore, we explicitly give the results and plot of a simulation
clearly showing that the claim of Iorio's paper that the $\dot
J_{2n}$ uncertainty may contribute to up a 45 $\%$ error error in
our measurement is clearly unsubstantiated. We then present a
rigorous proof that any ``imprint'' or ``memory'' effect of the
Lense-Thirring effect is completely negligible on the even zonal
harmonics produced using the GRACE satellites only and used on the
orbits of the LAGEOS satellites to measure the frame-dragging
effect. In this paper we do not discuss the problem of the
correlation of the Earth's even zonal harmonics since it only refers
to our previous, 1998, analysis with EGM96 and it will be the
subject of a different paper; nevertheless, we stress that in the
present analysis with EIGENGRACE02S the total error due to the
static Earth gravity field has been calculated by pessimistically
summing up the absolute values of the errors due to each Earth's
even zonal harmonic uncertainty, i.e., we have not used any
covariance matrix to calculate the total error but we have just
considered the worst possible contribution of each even zonal
harmonic uncertainty to the total error budget. We also present and
explain our past work on the technique of measuring the
Lense-Thirring effect using the LAGEOS nodes and give its main
references. Finally we discuss some other minor points and
misunderstandings of the paper by Iorio, including some obvious
mistakes contained both in this paper and in some other previous
papers of Iorio. In conclusion, the criticisms in Iorio's paper are
completely unfounded and misdirected: the uncertainties arising from
the possible variations of $\dot J_{2n}$ are fully accounted for in
the error budget that we have published.

\section{Error due to the $\dot J_{2n}$ in the 2004 measurement
of the Lense-Thirring effect}

In order to discuss the error analysis and the total error budget of
our measurement of the Lense-Thirring effect \cite{ciupav}, we first
stress that in the data reduction of our recent measurement of the
Lense-Thirring effect \cite{ciupav} we have used the value of $\dot
J_4 = -1.41\cdot10^{-11}$, adopted by GFZ  in the EIGEN-GRACE02S
Earth gravity model \cite{rei}, and we have fitted our combined
residuals with a secular trend only plus a number of periodical
terms. We can of course introduce $\dot J_4$ as a free parameter in
our fit (see below). In this case, together with the measurement of
the Lense-Thirring effect, we also  measure the effect of the
secular variations of $J_2$, $J_4$ and $J_6$ on the combination of
the nodal longitudes of the LAGEOS satellites; this is described by
a $\dot J_4^{Effective}$ \cite{eanes} in our combination, which
includes the effect of the secular variation of the higher even
zonal harmonics. In \cite{ciupavper} it is indeed reported an
effective value of $\dot J_4^{Effective} \cong - 1.5\cdot10^{-11}$
for the combination of the LAGEOS satellites nodes, which is
consistent with the EIGEN-GRACE02S model since, on our combination
of the nodal longitudes of the LAGEOS satellites, it just represents
a 6 $\%$ variation of the value given with EIGEN-GRACE02S and,
however, it includes the effect of any higher $\dot J_{2n}$, with
$2n \geq 4$; {\it this value is also fully consistent with our
published result of a Lense-Thirring drag equal to 99 $\%$ of the
general relativity prediction with uncertainty of 5 $\%$ to 10
$\%$}. It is easily seen, even by visual inspection, that our
combined residuals would clearly display any such quadratic term.
Indeed, in figure (1) we show the residuals obtained using the $\dot
J_4$ value given with EIGEN-GRACE02S that should be compared with a
simulation of the orbital residuals, shown in figure (5), obtained
using in the data reduction a strongly unrealistic value of $\dot
J_4$ corresponding to the value adopted in EIGEN-GRACE02S {\it plus}
12 times its published error and which produces a $\sim 45 \%$
variation of the secular trend as claimed by the author of
\cite{Iorio}! It is clear that only the first figure can be simply
described by a linear dependence.

\begin{figure}
\begin{center}
\includegraphics[scale=.8]{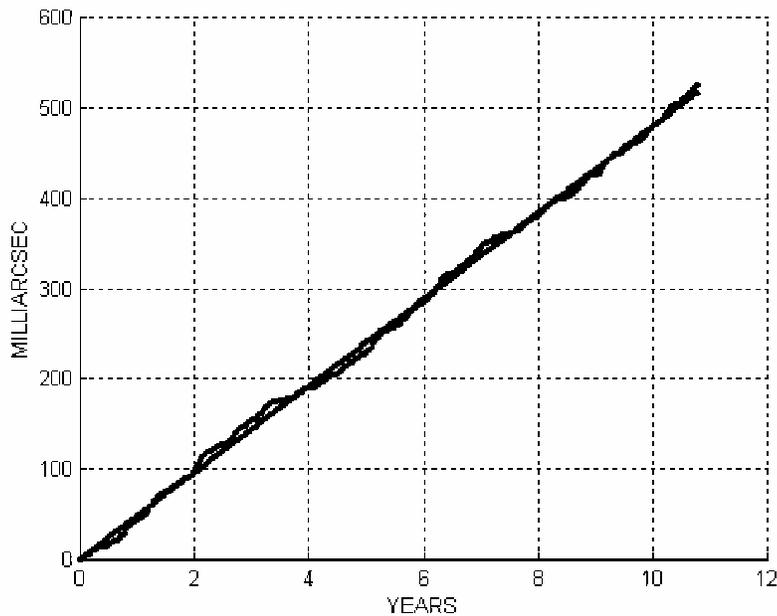}
\caption{Fit of the residuals of the nodes of LAGEOS and LAGEOS II,
using our combination and the Earth's model EIGEN-GRACE02S, with a
secular trend plus six periodical terms. The slope is \(\mu \simeq
0.99\) (Ciufolini and Pavlis 2004).}
\end{center}
\end{figure}

In the EIGEN-GRACE02S model \cite{rei}, obtained by the GRACE
mission only, the Earth gravity field was measured during the period
2002-2003. Corrections due to $\dot J_2$ and $\dot J_4$ were then
applied to this 2002-2003 measurement in order to obtain a gravity
field model antecedent to 2002-2003. These values of $\dot J_2$ and
$\dot J_4$, used by the GFZ team, are $\dot J_2 = -
2.6\cdot10^{-11}$ and $\dot J_4 = - 1.41\cdot10^{-11}$ and they were
measured on the basis of completely independent 30-year observations
before 2002.

{\it Let us describe the result of the orbital analyses using the
orbital estimator GEODYN with and without a contribution of $\dot
J_4 = - 1.41\cdot10^{-11}$, and the result of a simulation with a
$\dot J_4$ equal to 611 $\%$ of the value given in EIGEN-GRACE02S,
i.e. $\dot J_4 \simeq - 8.61\cdot10^{-11}$.}

First, we stress again that in the case of {\it not} applying the
$\dot J_4$ EIGEN-GRACE02S correction to the orbital analysis, its
effect can be clearly identified (but nevertheless completely fitted
for, see below) even by visual inspection since it shows up as a
hump in the combined residuals, whereas in our fit of \cite{ciupav}
shown in Fig. 1 the absence of any such quadratic effect is obvious.
Then, since the effect of the time variation $\dot J_4$ shows up as
a quadratic effect in the cumulative nodal longitude of the LAGEOS
satellites, the combined residuals of LAGEOS and LAGEOS 2 may be
fitted with a quadratic curve, together with a straight line and
with the main periodic terms. Thus, by fitting the raw residuals
obtained {\it without} any $\dot J_4$ in a $test-analysis$, see Fig.
2, we measured a $\dot J_4^{Effective} \cong - 1.5\cdot10^{-11}$,
which includes the effect of $\dot J_6$ and of higher even zonal
harmonics on the combination of the LAGEOS nodes. On other hand, by
fitting the combined residuals obtained with the EIGEN-GRACE02S
correction of $\dot{J_4}= - 1.41\cdot10^{-11}$, we measured a
$\delta \dot J_4^{Effective}$ of less than $- 0.1\cdot10^{-11}$, see
Fig. 3, and this fitted difference is in complete agreement with the
previous case. In all the cases of inclusion of an anomalous
$\dot{J_4}$ in the data analysis, we obtained a measured value of
$\mu \cong 0.98$, see Figures 3 and 4.

\begin{figure}
\begin{center}
\includegraphics[scale=.8]{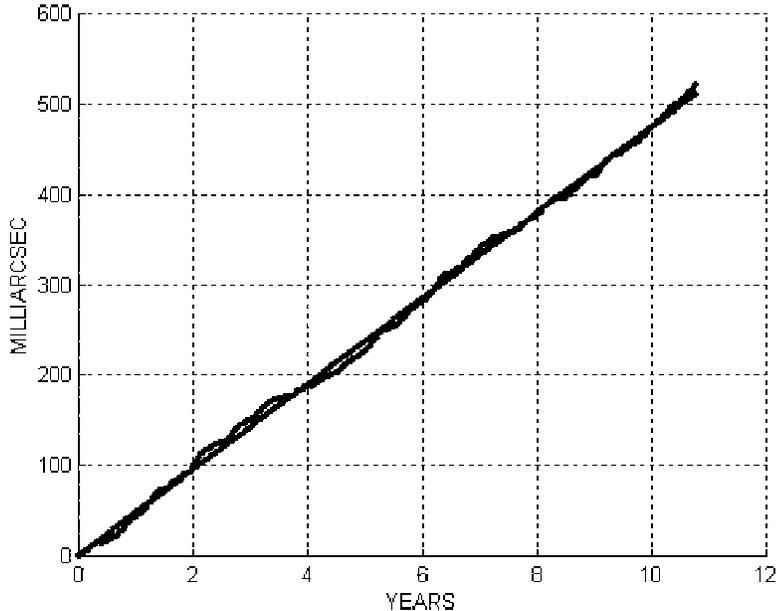}
\caption{Fit of the residuals of the nodes of LAGEOS and LAGEOS II,
using our combination and the Earth's model EIGEN-GRACE02S, with a
secular trend plus six periodical terms {\it plus a quadratic}. The
residuals have been obtained {\it without} inclusion of any $\dot
J_4$ value in the orbital analysis, i.e., using $\dot J_4 = 0$. The
Lense-Thirring slope is \(\mu \simeq 0.98\).} 
\end{center}
\end{figure}

\begin{figure}
\begin{center}
\includegraphics[scale=.8]{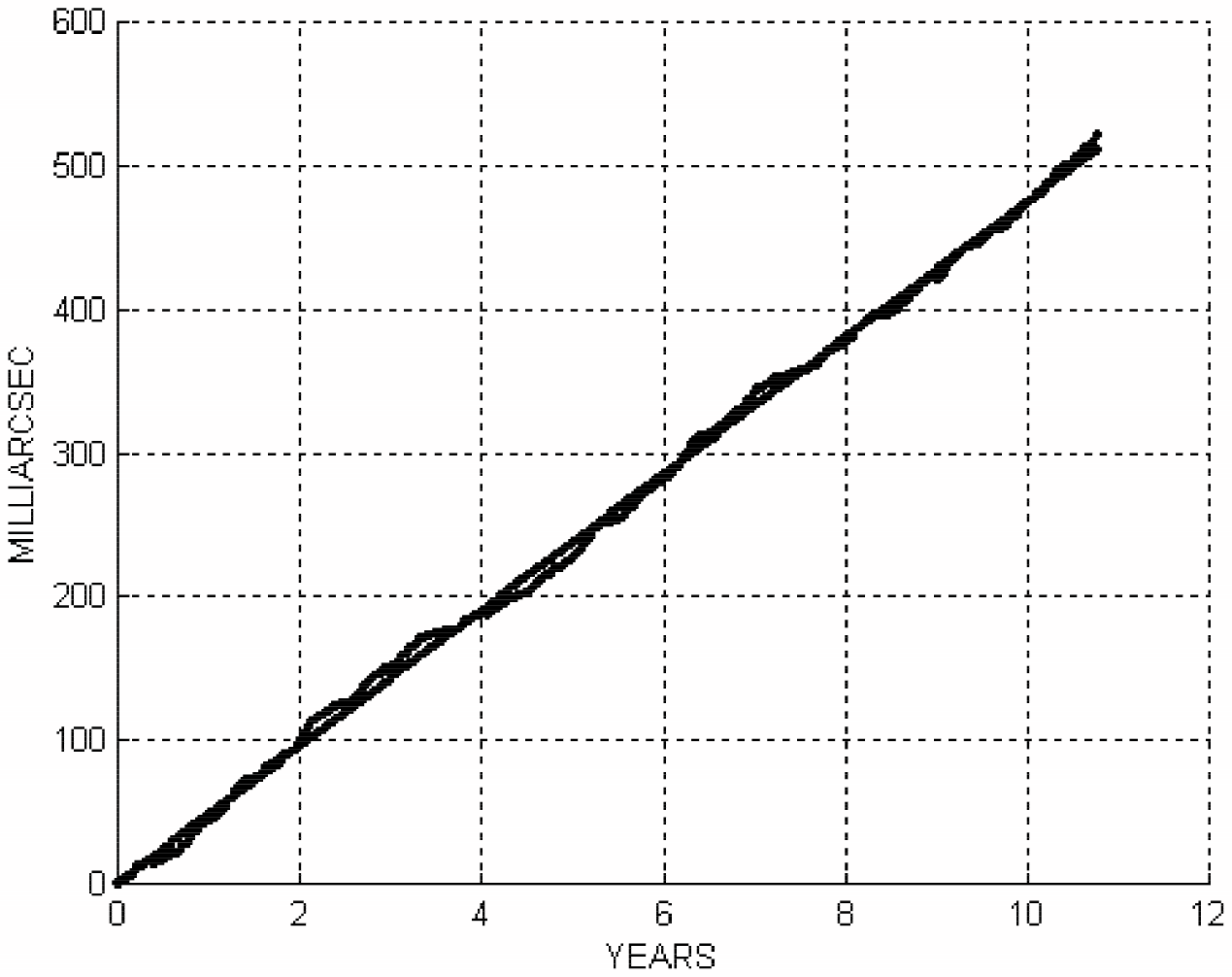}
\caption{Fit of the residuals of the nodes of LAGEOS and LAGEOS II,
using our combination and the Earth's model EIGEN-GRACE02S, with a
secular trend plus six periodical terms {\it plus a quadratic}. The
residuals have been obtained using the $\dot J_4$ value given in the
EIGEN-GRACE02S model. The Lense-Thirring slope is \(\mu \simeq
0.98\).} 
\end{center}
\end{figure}

\begin{figure}
\begin{center}
\includegraphics[scale=.8]{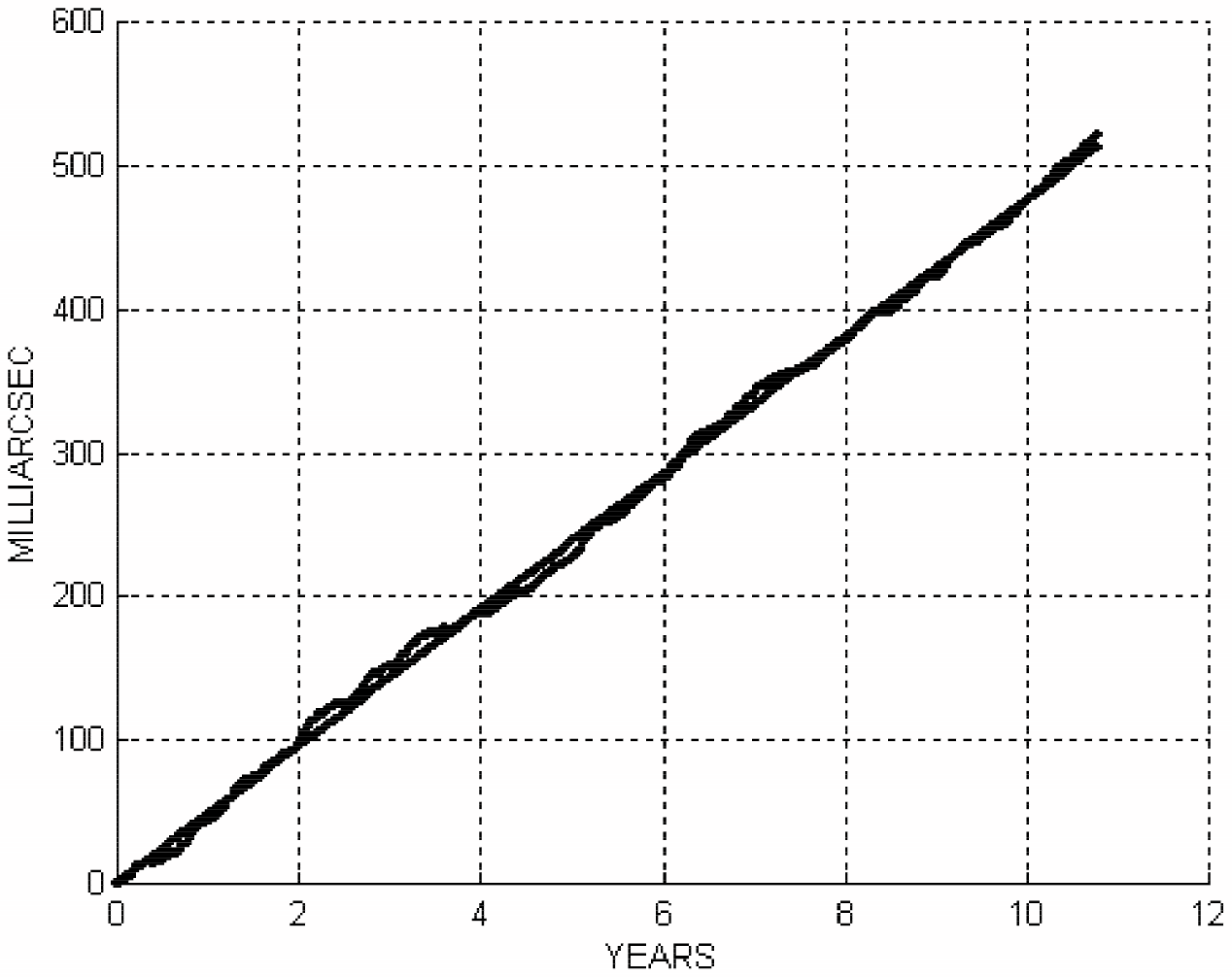}
\caption{Fit of the residuals of the nodes of LAGEOS and LAGEOS II,
using our combination and the Earth's model EIGEN-GRACE02S, with a
secular trend plus six periodical terms {\it plus a quadratic}. The
residuals have been obtained using a $\dot J_4$ value equal to 611
$\%$ of the $\dot J_4$ value given in the EIGEN-GRACE02S model. The
Lense-Thirring slope is \(\mu \simeq 0.98\).} 
\end{center}
\end{figure}

Therefore, the small value of the unmodelled quadratic effects in
our nodal combination due to the unmodelled $\dot J_{2n}$ effects
(with $2n \geq 4$) corresponds to a change in the measured value of
the Lense-Thirring effect of about 1 $\%$ only. As a {\it
consistency test}, using the value $\dot J_4^{Effective} = -
1.5\cdot10^{-11}$, obtained from fitting the combined residuals
(which is only about 6 $\%$ larger than the value $\dot J_4 = -
1.41\cdot10^{-11}$ given in the EIGEN-GRACE02S model), we have again
generated the residuals and fit these with a straight line only plus
the main periodic terms. It turned out that the change of the
measured value of frame-dragging was about 1 $\%$ only with respect
to the case of using $\dot J_4 - 1.41\cdot10^{-11}$ and fitting the
residuals with a straight line plus the main periodic terms
\cite{ciupav}, i.e., this test resulted in a measured value of $\mu
\cong 0.98$. The result of this test agrees with the case of fitting
the combined orbital residuals with a straight line plus a quadratic
curve, indeed in {\it all cases} we got $\mu \cong 0.98$, see
Figures 2, 3 and 4. As a third case, using in the data reduction a
highly unrealistic value of $\dot J_4$ equal to the value adopted in
EIGEN-GRACE02S {\it plus} 12 times its published error, i.e., a
$\dot J_4$ equal to about 611 $\%$ of the value given in
EIGEN-GRACE02S, that is $\dot J_4 = 6.11 \times (-1.41\cdot10^{-11})
\simeq -8.61\cdot10^{-11}$, by fitting the residuals with a straight
line plus the main periodic terms plus a parabola, we again obtained
the same measured value of frame-dragging, i.e., $\mu \cong 0.98$,
see Fig. 4. Therefore, also in this case the measured value of
frame-dragging differs by about 1 $\%$ {\it only} with respect to
the case of using $\dot J_4 = - 1.41\cdot10^{-11}$ without fitting a
quadratic curve, i.e., the result reported in \cite{ciupav} and in
Fig. (1).

These analyses clearly show that even in the cases of - 100 $\%$ and
+ 511 $\%$ variations between the value of $\dot J_4$ of Mother
Nature and the value used in the orbital estimation of
\cite{ciupav}, it is possible to fit for the $\dot J_4^{Effective}$
effect and get back from the fit most of the simulated variation in
$\dot J_4$ but especially it shows that the measured value of the
Lense-Thirring effect can only be affected at the 1 $\%$ level by
the $\dot J_{2n}$ uncertainties. Indeed, by fitting with a parabola
the combined orbital residuals in these three cases corresponding to
different values of $\dot J_4$ (zero, the EIGEN-GRACE02S value, and
611 $\%$ of this EIGEN-GRACE02S value), we always obtained $\mu
\cong 0.98$ (Figures 2, 3 and 4, whereas in \cite{ciupav} we
obtained, $without \, fitting \, any \, parabola$, $\mu \cong 0.99$.

 In conclusion this 1 $\%$ variation
(obtained by fitting the residuals including a parabola) with
respect to the value of $\mu \cong 0.99$ measured in \cite{ciupav}
(without any parabola in the fit) gives the estimated error due to
the $\dot J_{2n}$ uncertainties in our measurement of $\mu$, as
reported in \cite{ciupav}. In \cite{Iorio}, it is claimed that the 5
$\%$ to 10 $\%$ error budget of our recent measurement has been
strongly underestimated because of the $\dot J_4$ and $\dot J_6$
errors; indeed, in this paper it is claimed a possible error as
large as 45 $\%$  due to the $\dot J_{2n}$ uncertainties. However,
this statement is simply nonsense based on what we explained in the
previous paragraph and only shows a lack of understanding of both
the real effect of the $\dot J_{2n}$, with $2n \geq 4$, on the
combination of the nodes of LAGEOS and LAGEOS 2 and of the technique
of the least-squares fit.

As already pointed out, it is critical to stress that anyone can
immediately rule out a large error of this type even by visual
inspection of Fig. 1, which is the fit of the combined orbital
residuals of \cite{ciupav}. A change, for example, of 45 $\%$ in the
fit due to the $\dot J_{2n}$ effects (as claimed in \cite{Iorio})
would correspond to a superposition of a gigantic hump, with a
height of several hundreds of milliarcsec, to our {\it raw}
residuals that instead clearly show a straight line trend with a
rate of 47.9 milliarcsec (compare Fig. 1 and Fig 5)!

\begin{figure}
\begin{center}
\includegraphics[scale=.8]{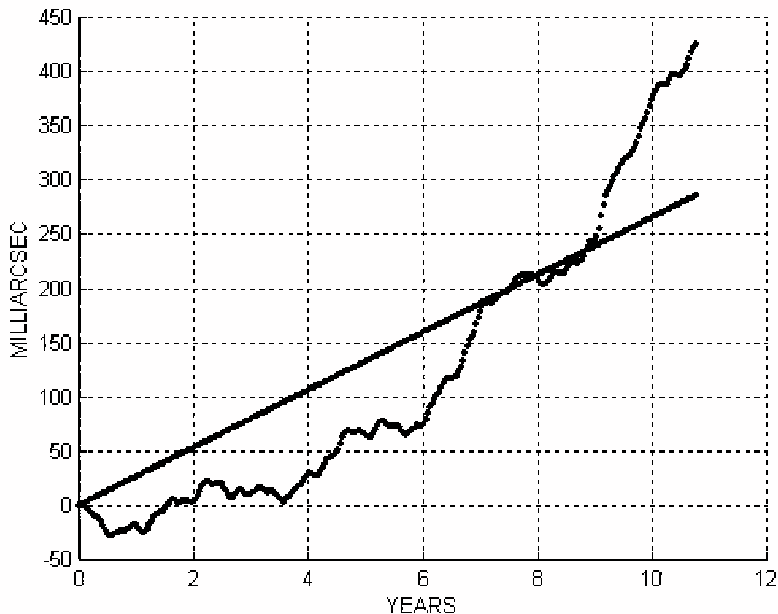}
\caption{Simulation of our combined residuals of the nodes of LAGEOS
and LAGEOS II using our combination. The residuals have been
obtained using a highly unrealistic value of $\dot J_4$ equal to 611
$\%$ of the $\dot J_4$ value given in the EIGEN-GRACE02S model with
x-intercept in 2000. The Lense-Thirring slope, fitting only for a
secular trend, is \(\mu \simeq 0.55\), i.e., it has a change of
about 45 $\%$ with respect to the fit of Figure 2. The quadratic
effect in the combined residuals due to $\dot J_4$ can be clearly
identified even by visual inspection. This plot should be compared
with Fig. 1 (the result of (Ciufolini and Pavlis 2004)) and with
Fig. 4 where these {\it simulated} combined residuals have been
fitted by including a quadratic term in order to fit for the $\dot
J_4^{Effective}$ effect.}
\end{center}
\end{figure}

Indeed, in Fig. 5 we have shown the result of a {\it simulation}
using, as an example, a value of $\dot J_4 = - 8.61\cdot10^{-11}$,
corresponding to the value given in the EIGEN-GRACE02S model $plus$
12 times the error of $\dot J_{4}$ given in \cite{cox} (see below),
for example with x-intercept in the year 2000. Not only, in this
case, to get a 45 $\%$ error we would need an error in $\dot J_{4}$
that is 1200 $\%$ of what estimated in ref. \cite {cox}, but any
such error would produce a huge deformation of the final residuals
that can be easily identified even by visual inspection of the
combined residuals and, however, completely fitted for using a
quadratic as shown in Fig. 4 and as previously explained.
 A 45 $\%$ error in the measured value of $\mu$ due to a
superimposed quadratic effect looks as a clear nonsense! In other
words, by comparing Fig. (5) with Fig. (1) (corresponding to our
measurement \cite{ciupav}), it is clear that what is missing and
misunderstood in \cite{Iorio} is that the only {\it relevant}
secular effect that can mimic the Lense-Thirring effect is due to
the errors in the values of the static $J_{2n}$ coefficients used in
our analysis, whereas the $\dot J_{2n}$ will show up as quadratic
effects and therefore large values of $\dot J_{2n}$ can be {\it
clearly identified} over our long period of 11 years and can be
fitted for, using a quadratic curve. For example, in the single,
i.e., $\it not$ combined, residuals of the node of LAGEOS and of the
node of LAGEOS 2 of Fig. 6 and 7 is clearly observed a long term
variation in the trend that, since it disappears in the nodes
combination of Fig. 1, can be identified as an anomalous increase in
the Earth quadrupole moment, corresponding to the effect observed by
Cox and Chao \cite{cox}.

Finally, since the measurements of $\dot J_{2n}$ and $J_{2n}$ are
completely independent (as previously explained), the corresponding
errors are independent and this allowed us to take the RSS of these
errors. Nevertheless, even by just adding up these errors we would
get an error budget of 6.6 $\%$, well within our quoted range of 5
$\%$ to 10 $\%$.

\begin{figure}
\begin{center}
\includegraphics[scale=.8]{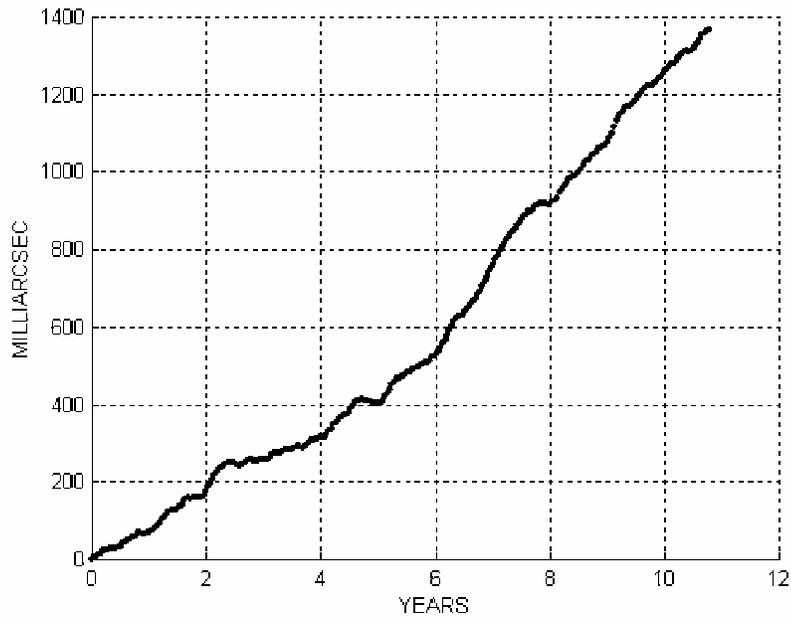}
\caption{Residuals of the nodal longitude of LAGEOS using
EIGEN-GRACE02S.}
\end{center}
\end{figure}

\begin{figure}
\begin{center}
\includegraphics[scale=.8]{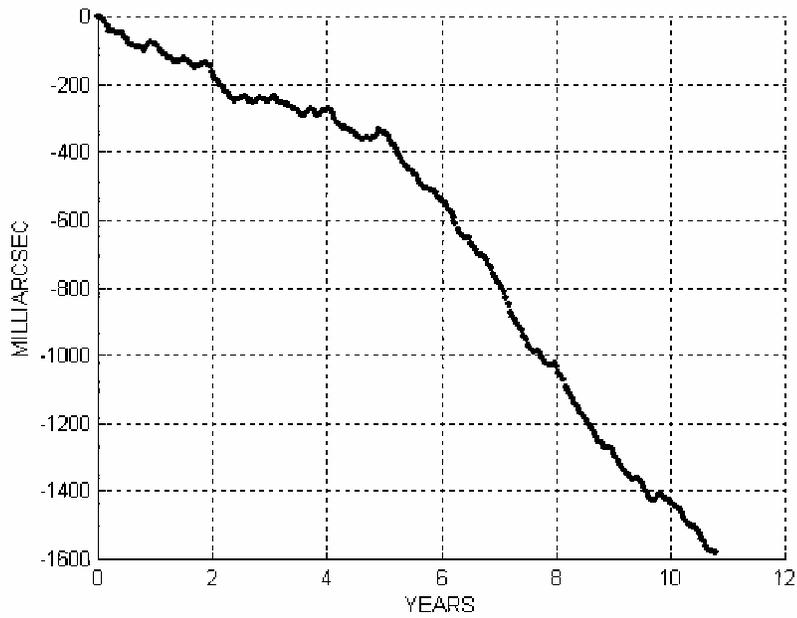}
\caption{Residuals of the nodal longitude of LAGEOS 2 using
EIGEN-GRACE02S.}
\end{center}
\end{figure}

Another way to look at the interconnection between the
Lense-Thirring effect and the value of $\dot J_{4}$ is through a
$\chi^2$ plot, as shown in Fig. 8. The origin of this graph
corresponds to the best fit of our combined residuals by including a
parabola in the fit (i.e., it corresponds to $\dot J_4^{Effective}
\cong - 1.41\cdot10^{-11} - 0.1\cdot10^{-11} = - 1.51\cdot10^{-11}$
and Lense-Thirring effect $= 98 \%$ of the general relativity
prediction) and the ellipse represents the confidence level
\cite{press} for a 99 $\%$ probability. It has to be stressed that
this ellipse of confidence is tentative in the sense that we have
assumed an {\it average} tentative standard deviation for each
combined residual; the full analysis using the covariance matrix
obtained by the GEODYN data reduction  is presented in
\cite{ciupavnext} . In this figure we have also included the value
of $\dot J_{4}$ adopted in the EIGEN-GRACE02S model \cite{cox}. The
two parameters are indeed correlated, but this figure shows that the
maximum uncertainty in the Lense-Thirring parameter arising from the
possible variation of $\dot J_{4}$ is just $\sim \pm 0.75 \%$ of its
value. The complete error analysis published in \cite{ciupav} and
\cite{ciupavper} quotes a larger total uncertainty between 5 $\%$
and 10 $\%$ since it also takes into account all the other error
sources.

\begin{figure}
\begin{center}
\includegraphics[scale=.8]{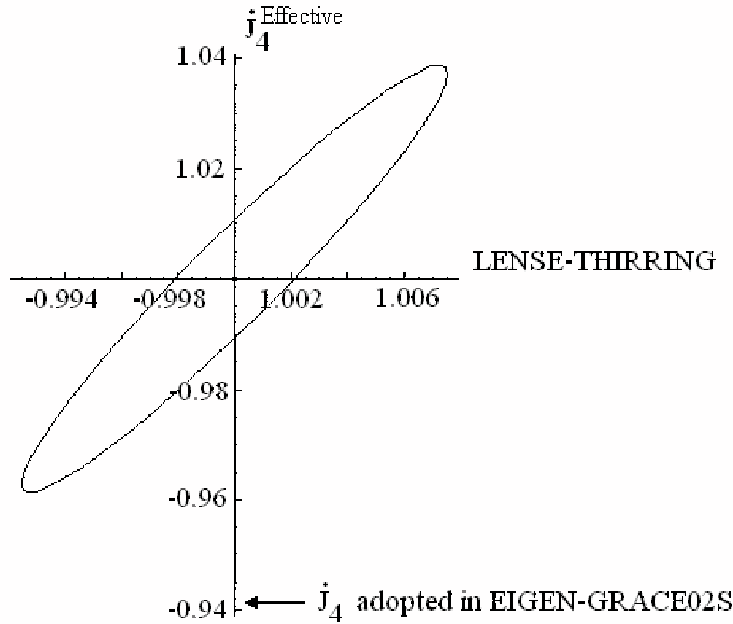}
\caption{Confidence Ellipse. On the $x$ axis is the Lense-Thirring
effect; the origin, at $x=1$, corresponds to the fitted value of
frame-dragging. On the $y$ axis is $\dot J_4^{Effective}$; the
origin, at $y=1$, corresponds to the fitted value $\dot
J_4^{Effective} \simeq - 1.51\cdot10^{-11}$. This confidence ellipse
is tentative since we have assumed an average tentative standard
deviation for each combined residual. This figure shows that the maximum uncertainty in the measured Lense-Thirring parameter arising from the possible variation of $\dot J_{4}$ is just $\sim 0.75 \%$}
\label{fig:}
\end{center}
\end{figure}


Let us estimate with a different method the maximum conceivable
error due to the the $\dot J_{2n}$ in our measurement of the
Lense-Thirring effect. First {\it we stress that our error budget
relies on the validity of the EIGEN-GRACE02S model and of the GRACE
mission. Questioning our error budget is basically equivalent to
question the validity of the GRACE mission itself.}

As already pointed out, the Earth model EIGEN-GRACE02S was produced
by GRACE measurements {\it only}, during the period 2002-2003,
whereas the {\it independent} evaluation of $\dot J_4$ was produced
on the basis of completely independent 30-year observations {\it
before} 2002. The only constraint (based on the validity of the
GRACE measurements) is that the $\dot J_4$ correction applied to
$J_4$, {\it must} of course produce in 2002-2003 the same value of
$J_4$ that was measured by GRACE in 2002-2003, at least within the
EIGEN-GRACE02S uncertainties in $J_{4}$. Running then the orbital
estimator GEODYN  with or without a contribution corresponding to
the estimated error in $\dot J_4$ of $ \pm 0.6 \cdot 10^{-11}$
\cite{cox} and {\it by fitting the residuals without a parabola}
resulted in a change of the corresponding measured value of
frame-dragging of $\pm$ 7.29 $\%$. This was verified in {\it two}
ways: using GEODYN and an independent program written specifically
to evaluate the $\dot J_4$ effect. By changing the observational
period the measured value of frame-dragging would only slightly
change (see the slope of Fig.1 over subsets of the total
observational period of 11 years) and the corresponding error would
{\it consistently} change too, increasing over periods farther away
from 2002-2003. Nevertheless, over our observational period, the
maximum $\dot J_4$ error was 7.29 $\%$. Similarly, {\it by fitting
the residuals without a parabola}, an error of $0.5 \cdot 10^{-11}$
in $\dot J_6$ would produce a 2.98 $\%$ error over our observational
period. Then, by adding these errors to the error due to tides we
get a maximum error of 11 $\%$ and finally considering all the other
independent error sources (see the Supplementary Discussion of ref.
\cite{ciupav}) we get a total uncertainty of 12 $\%$ in our
measurement of the Lense-Thirring effect, in substantial agreement
with the maximum error of {\it about} 10 $\%$ reported in
\cite{ciupav}. Nevertheless, it is critical to perform this worst
case error analysis (without fitting our residuals with a quadratic)
by using the most recent 2004 measurements of $\dot J_4$ and $\dot
J_6$ \cite{2004AGUFM.G31C0801C}, see also section 2. Recently, a
solution for the time variation of the even zonal harmonics up to
degree 6 was determined using Satellite Laser Ranging data from 8
geodetic satellites over a time period of 28 years from 1976 to
2003. The secular zonal variations for the degree up to 6 have been
determined using the existing long term Satellite Laser Ranging data
from multiple satellites, including Starlette, Ajisai, Stella,
Lageos I and II, Etalon I and II, and BEC using both long and short
arc orbital analysis technique. The results of
\cite{2004AGUFM.G31C0801C} are $\dot J_4 = - 1.99\cdot10^{-11}$ and
$\dot J_6 = 0.02\cdot10^{-11}$. Then, we have used these values of
$\dot J_4$ and $\dot J_6$ and the orbital estimator GEODYN,
according to the worst case analysis just described (without any fit
for a quadratic term), to evaluate the impact of these different
values of $\dot J_4$ and $\dot J_6$ in our measurement of the
Lense-Thirring effect. We have then found a total difference of
about 6.9 $\%$ in our measurement of the Lense-Thirring effect due
to these new values of $\dot J_4$ and $\dot J_6$ (a precise
statistical analysis based on the scatter of the various
determinations of the $\dot J_{2n}$ is presented in a detailed paper
in preparation \cite{ciupavnext}). Then, by adding this error to the
error due to tides we get a maximum error of less than 7.9 $\%$ from
time variations of the Earth gravity field and finally considering
all the other independent error sources we get a total uncertainty
of 9.3 $\%$ in our measurement of the Lense-Thirring effect, {\it in
full agreement with the maximum error of {\it about} 10 $\%$
reported in \cite{ciupav}}.

However, as explained above, we stress that the effect of such an
error in $\dot J_4^{Effective}$ is {\it not} observed in our
combined residuals, nor is of course observed a larger error of this
type \cite{Iorio} that would in fact produce in our combined
residuals a huge parabola superimposed to the frame-dragging trend
(see, for example, Fig. 5). Indeed, as previously explained, by
fitting our raw residuals including a quadratic curve we have that
the total effect of the $\dot J_{2n}$, with ${2n} \geq 4$, on the
combined node residuals corresponds to a total error in our
determination of frame-dragging of about 1 $\%$ only.

It is finally worth to stress once again that the measured value of
the Lense-Thirring effect, $\mu \cong 0.98$, obtained by fitting
with a straight line and six periodic terms {\it plus} a parabola
the raw combined residuals of \cite{ciupav}, fit shown in Fig. 2,
fully agrees with the measured value of the Lense-Thirring effect,
$\mu \cong 0.98$, obtained by fitting with a straight line and six
periodic terms {\it plus} a parabola the raw combined residuals
simulated over our observational period of 11 years using GEODYN
{\it without} inclusion of any $\dot J_4$, i.e. with $\dot J_4 = 0$
(Fig. 3), and with a highly unrealistic value of $\dot J_4 = 611 \,
\% \, {\dot J_4}^{EIGEN-GRACE02S}$ (Fig. 4).

In conclusion, the 5 $\%$ uncertainty in our  measurement of the
Lense-Thirring effect is fully confirmed in spite of the serious
misunderstandings of our analysis in \cite{Iorio}. Indeed, a number
of further critical remarks are reported in \cite{Iorio} that show
that his author did not carefully and seriously read our papers but
only superficially read them and deeply misunderstood them.

In \cite{iorio2} is written ``The value $\dot J_4^{eff} \cong -
1.5\cdot10^{-11}$ year$^{-1}$ measured with the combination of Eq.
(5) is affected not only by $\dot J_6$ and the other higher degree
even zonal harmonics but, more importantly, by the Lense–Thirring
signature itself. Indeed, the combination of Eq. (5) is designed in
order to only disentangle $\dot J_6$ and the Lense–Thirring effect.
So, it is not admissible to use the so obtained $\dot J_4^{eff}$,
which is coupled by construction to the Lense–Thirring effect, in
order to reliably and correctly measure the Lense–Thirring effect
itself'' This statement is {\it completely false} because the
Lense-Thirring effect was measured in \cite{ciupav}, as very clearly
written in \cite{ciupav}, by only fitting with a straight line (plus
the main periodical terms) the residuals produced with the standard
value of $\dot J_4$ adopted in the EIGEN-GRACE02S model: no
effective $\dot J_4$ and no quadratic fit were used in order to
obtain our result in \cite{ciupav}. The quadratic fit, together with
a straight line and harmonics terms, was only used as a "tool", to
evaluate the strength of any unmodelled $\dot J_4$ effect (as above
explained in details). Furthermore, as clearly explained above, the
statement ``the so obtained $\dot J_4^{eff}$, which is coupled by
construction to the Lense–Thirring effect'' is completely void of
any possible meaning.

Furthermore, this paper \cite{iorio2} is not $properly$ referring to
all the fundamental papers published since 1984 on the use of the
nodes of the LAGEOS satellites in order to measure the
Lense-Thirring effect (see section 4).

In the conclusion of  \cite{iorio2} is written: ``For a given
observational time span and a given background reference Earth
gravity model, two time series of the combined residuals, built up
with and without the aforementioned $\dot J_l$ GTR-free values in
the background reference models, it should be analyzed. The
difference between the so obtained $\mu$ parameters can, then, be
evaluated.'' Again, in an official rebuttal letter of Ciufolini and
Pavlis, dated November 2004 \cite{ciupav2}, to answer one of the
papers of criticism sent by Iorio for publication (rejected by the
referees), the method to evaluate the impact of the $\dot J_{2n}$
uncertainties was clearly written and explained: `` ... Running then
the orbital estimator GEODYN with or without a contribution
corresponding to the estimated error in $\dot J_4$ of $0.6 \cdot
10^{-11}$ \cite{cox} resulted in a change of the corresponding
measured value of frame-dragging by 7.29 $\%$. ...''. This is
exactly one of the methods of error analysis reported here.

However, the most {\it profound misunderstanding} in \cite{iorio2}
is the following ``... If one is interested in extracting from such
a signal a linear trend, it is not possible also to fit a parabolic
noise signal without corrupting or distorting in some way the
genuine linear trend of interest. Exactly in the same way, it is not
possible to fit and remove the noise linear signal induced by $J_4$,
$J_6$, . . .  It is only possible to evaluate, as more accurately
and realistically as possible, their impact on the measurement of
the linear trend.''. In a previous similar paper of the same author
\cite{iorio2} is written: ``The time series of the combined
residuals of the nodes of LAGEOS and LAGEOS II consists of the
Lense-Thirring linear trend as predicted by GTR, which has, in fact,
to be measured, the linear trend induced by the static part of the
uncancelled even zonal harmonics of the geopotential $J_4$, $J_6$
..., the parabolic signal induced by $\dot J_4$, $\dot J_6$ and the
time-varying harmonic perturbations of gravitational and non
gravitational origin. If one is interested in extracting from such a
signal a linear trend, modelled as $a_0 + a_1 t$, it is not possible
also to fit a parabolic noise signal, modelled as $b_0 + b_1 t + b_2
t^2$, without corrupting or distorting in some way the genuine
linear trend of interest. ...''.

However, in the above mentioned official letter of Ciufolini and
Pavlis to Iorio, dated November 2004 \cite{ciupav2}, it was clearly
written: ``First, our error budget relies on the validity of the
EIGEN-GRACE02S model and of the GRACE mission. Questioning our error
budget is equivalent to questioning the validity of the GRACE
mission itself. We stress that the Earth model EIGEN-GRACE02S was
produced by GRACE measurements ONLY, during the period 2002-2003,
whereas the INDEPENDENT evaluation of $J_4$-dot was produced on the
basis of completely independent 30-year observations BEFORE 2002.
The only constraint (based on the validity of the GRACE
measurements) is that the $J_4$-dot correction applied to $J_4$,
MUST of course produce in 2002-2003 the same value of $J_4$ as that
was measured by GRACE in 2002-2003.''

This  means that one is not free to fit for a trend $b_1 t$ due to
$\dot J_4$ in addition to the trend $a_1 t$ but, on the basis of the
2002-2003 GRACE measurement of $J_4$, any change $\dot J_4$, when
applied to the EIGEN-GRACE02S static model, must of course be such
to agree in 2002-2003 with the value of $J_4$ measured by GRACE in
2002-2003 (at least within the small measurement uncertainty given
in the EIGEN-GRACE02S model; this small uncertainty has however a
negligible influence on our final result). This means that, since we
have to be consistent with the GRACE observation of $J_4$ in
2002-2003, the term $b_1$ must be: $b_1 \equiv - 2 b_2 N$, where N
is the number of years between the first year of application of the
$\dot J_4$ correction and the GRACE observations in 2002-2003.
Therefore, as we did, one only needs to fit for $b_2$ and get $b_1$
from it, or alternatively (as we also did) fit for $b_2 (t^2 - 2 N
t)$ in addition to the $a_1 t$ trend. So our test fit is just a
polynomial fit (with or without the periodic terms). {\it In
conclusion, in the two sections reported above, the author of
\cite{Iorio} is explicitly stating that one cannot fit some
observational data with a polynomial of degree three!!}

\section{On the value of $\dot{J_4}$}

Long period and secular variations in the zonal harmonics of the
spherical harmonic expansion describing Earth's gravitational field
are generally attributed to ``post-glacial rebound'' and ice-sheet
mass changes
\cite{1992GeoJI.108....1T,1993JGR....98.4579I,1993JGR....98.4509M},
and reflect changes in Earth's viscous lower mantle
\cite{1993JGR....98.4579I}. These changes hold information about the
dynamics of Earth and its anelastic response to tidal forcing.
Estimation of the secular rates for the zonal harmonics therefore
has been a high priority research topic of space geodesy for many
years now, with a considerable number of independent results in the
literature
\cite{1983Natur.303..757Y,1984JGR....89.1077R,1989GeoRL..16..393C,1991GeoRL..18.1739G,1993GeoRL..20..595N,1993Cheng,1996Cazenave,1996Nerem,1996Eanes,1997JGR...10222377C,1999Devoti,2000Cox,2002Pavlis,2004JGRB..10909402C,2004AGUFM.G31C0801C}.
The majority of these results were obtained from the analysis of
long time intervals spanning several decades of data obtained from
several satellites including the two LAGEOS, Starlette, Ajisai,
Stella, BE-C and the two ETALONs. Due to the limited number of
satellite orbit configurations, only a handful of the very lowest
degree zonal terms' rates can be ``observed'' \cite{eanes}, reaching
at best, up to degree six. The estimation of the zonal rates is in
most cases based on perturbation analysis of very long arcs (several
years), but not always necessarily so. In particular, the last four
studies cited above involve short-arc techniques, estimating the
average ``static'' value of the zonal harmonics, and deriving the
rate from a subsequent analysis of the resulting time series.
Furthermore, in most cases, the investigators used their static
gravitational model of preference, implying that the adopted epoch
values for $J_4$ were different amongst studies. These differences
are well documented in the literature dealing with the static models
JGM-3, EGM96, GRIM5-S1, TEG-4, EIGEN-GRACE02S, and GGM01S. It is
then remarkable that from a plethora of approaches, analyzed data
sets, and diverse analysis s/w packages used, the resulting estimate
for the secular rate of these harmonics, e.g. \(\dot{J}_4\), which
is of interest here, differs by no more than 100\% from the commonly
accepted value \cite{mc}, the one that was also used in the
development of EIGEN-GRACE02S.

As it is discussed in \cite{1997JGR...10222377C}, there is
considerable correlation between the zonal rate estimates
(\(\rho_{\dot{J_2},\dot{J_4}}=-0.75,
\rho_{\dot{J_2},\dot{J_6}}=+0.76,
\rho_{\dot{J_4},\dot{J_6}}=-0.86\)), but there is no significant
correlation between the static values of the harmonics and their
rates. This independence of errors between the static values'
estimates and the associated rates is further demonstrated in
\cite{2004AGUFM.G31C0801C}, where monthly estimates of the \(J_2\)
term obtained from GRACE data and those obtained entirely
independently from SLR data analysis are compared. The agreement is
remarkable, even though the GRACE data analysis has not reached yet
its definitive level. The results from \cite{2004AGUFM.G31C0801C}
suggest that the current best estimate for \(\dot{J_4}\), with all
other zonal rates up to \(\dot{J_6}\) simultaneously estimated, is
\(-1.99 \times 10^{-11}\)/y. This value is only \(0.58 \times
10^{-11}\)/y off the currently accepted value, adopted in developing
EIGEN-GRACE02S, or in terms of a percentage, it is off by only
-41\%. This possible range of error is well within our \(\mu\) error
margin (5-10\%), since our simulation of a 611\% error in
\(\dot{J_4}\) resulted only in a corresponding 1\% error in the
fitted value of \(\mu\).

A careful collection of all the published values of \(\dot{J_4}\) in
the literature and subsequent statistical analysis, points to the
same conclusion. Since the published estimates can be in most cases
considered independent as we explained above, we can perform a
simple statistical analysis of these values and their variance in
comparison to the quoted error estimates of these values. The
population is not large by any statistical standards, but that is
what is available. There are two estimates, \(+0.2 \times 10^{-11}\)
\cite{1996Nerem} and \(+0.3 \times 10^{-11}\)
\cite{1989GeoRL..16..393C}, which are clearly outliers and the
reasons are explained in \cite{1997JGR...10222377C}, so our
statistical tests include cases where both of these or either one
separately are eliminated from the population of estimates. The
results in terms of a mean \(\dot{J_4}\) and its associated scatter
are shown in Table \ref{tab:J4dot}.

 \begin{table}
\caption{Statistical significance of the variations between
individual \(\dot{J_4}\) estimates.} \label{tab:J4dot}
\begin{center}
\begin{tabular}{lccc}
\noalign{\hrule height 1.0pt} Case Description & Mean value of
\(\dot{J_4}\) & Scatter of \(\dot{J_4}\) & References\\
& (\(\times 10^{-11}\)/y) & (\(\times 10^{-11}\)/y)\\\hline A. Six
estimates\\ $\;\;$ from Table 5\\ $\;\;$ (ref. 1) and\\ $\;\;$ the new estimate\\
$\;\;$ from
(ref. 2) & -0.70 & \(\pm\)0.8 & (1): \cite{1997JGR...10222377C}\\
& & & (2): \cite{2004AGUFM.G31C0801C}\\
B. As in A,\\ $\;\;$ but without\\ $\;\;$ the two outliers\\ $\;\;$ of +0.2 and\\
$\;\;$ \(+0.3 \times 10^{-11}\)/y & -1.02 & \(\pm\)0.6 & As in (A).\\
C. As in A,\\ $\;\;$ removing only\\ $\;\;$ the estimate of\\
$\;\;$ \(+0.3 \times 10^{-11}\)/y & -0.84 & \(\pm\)0.74 & As in (A).\\
D. As in A,\\ $\;\;$ removing only\\ $\;\;$ the estimate of\\
$\;\;$ \(+0.2 \times 10^{-11}\)/y & -0.83 & \(\pm\) 0.76 & As in (A).\\
E. Nine estimates\\ $\;\;$ from Table 7\\ $\;\;$ of Ref. 1. & -1.697
&
\(\pm\)0.69 & (1): \cite{1997JGR...10222377C}\\
\noalign{\hrule height 1.0pt}
\end{tabular}
\end{center}
\end{table}

These tests point to one conclusion very clearly: the uncertainty in
our present knowledge of \(\dot{J_4}\) is in the neighborhood of
\(\pm 0.7 \times 10^{-11}\)/y and this quantity is only 50\% of the
\(\dot{J_4}\) value that we used in our Lense--Thirring measurement.
It implies that with a 67\% probability, the true value of
\(\dot{J_4}\) could be in the interval [-1.697 - 0.7 to -1.697 +
0.7] , that is [-2.4, -1.0], or if we increase the probability to
95\% confidence level, we then have the corresponding interval of
[-3.1, -0.3], or even at the 99\% level, \(\dot{J_4}\) will then lie
in the interval [-3.8, +0.4], all values in \(10^{-11}\)/y.

From all these cases, we see that we never reach the extreme valueon: \(-8.61 \times 10^{-11}\)/y,
although, even in that case, the final effect on the fitted value of
\(\mu\) was only 1\%!

In conclusion, from all the cases studied, there is no reason to
expect an error in \(\dot{J_4}\) of more than \(0.7 \times
10^{-11}\)/y (at 1-\(\sigma\) level of significance, i.e. 67\%
probability), and all statistically acceptable variations in the
estimate of \(\dot{J_4}\) indicate that there can be no more than
1\% error in the fitted value of \(\mu\), caused by this error in
\(\dot{J_4}\), as shown in section 1. The evaluation of the
\(\dot{J_4}\) error estimate on the basis of the various results
indicates also a perfect agreement with the quoted \(1 \times
10^{-11}\)/y error for the \(-1.41 \times 10^{-11}\)/y rate for
\(J_4\), which was adopted in developing EIGEN-GRACE02S.

\section{``Imprint'' or ``Memory'' of the Lense-Thirring Effect}

In regard to other conceivable error sources, a possible bias in our
measurement of \(\mu\) might be due to some ``imprint'' or
``memory'' of the a priori value of the Lense--Thirring effect used
in the determination of the gravity model [see: \cite{ciucqg97}].
The author of \cite{Iorio} is mentioning this potential source of
error in section 3.2.6 ``The a priori ``memory'' effect of the
Lense– Thirring signature on the adopted Earth gravity model on the
Earth gravity model used''. The Earth gravity field model
EIGEN-GRACE02S has been obtained using the observations of the GRACE
satellites {\it only} and by using a set of models and parameters to
describe the orbital perturbations, including some a priori,
theoretical, value of the Lense--Thirring effect, indeed
EIGEN-GRACE02S has been obtained without the Lense-Thirring effect
in the data analysis. The previous models EIGEN-2S and GGM01S have
been derived using the observations of the CHAMP and GRACE
satellites. Thus, the gravitational field model EIGEN-GRACE02S that
we used to measure \(\mu\) may contain some kind of ``imprint'' of
the a priori, theoretical, value of the Lense--Thirring effect
(i.e., zero in the EIGEN-GRACE02S case).

Nevertheless, in the present analysis we are only concerned about a
conceivable imprint on the Earth's even zonal harmonics \(J_{2n}\).
The Earth gravitational potential is measured by observation of the
rate of change of the GRACE inter-satellite distance \cite{rei},
i.e., the gravitational field is measured by observation of the
relative acceleration of two test particles (the equation linking
relative acceleration and gravitational field is, in general
relativity, the geodesic deviation equation \cite{ciuwhe}).
 However, the effect of the gravitomagnetic field on the acceleration
vector of a satellite \cite{ciuwhe}, decomposed in out--of--plane
direction, \(\bot\) (orthogonal to the satellite orbital plane);
along track, \(\hat{\bf{v}}\) (along the satellite velocity vector);
and radial direction, \(\hat{\bf{r}}\) (along \(\bf{r}\)), is given
by \(\bf{a} = \bf{v} \times \bf{H}\), where \({\bf H} \, = \, {{{2
\, [ {\bf J} \, - \, 3 \, ( {\bf J} \, \cdot \, \hat {\bf x} \, ) \,
\hat {\bf x}]} \over {\vert {\bf x} \vert^{3}}}}\) is the
gravitomagnetic field and \({\bf{v}} = v \bf\hat{v}\). For a polar
satellite with a circular orbit we then have \(a^{L-T}_\bot \sim
\sin \phi\), where \(\phi\) is the latitude,
\(a^{L-T}_{\hat{\bf{r}}} = 0\) and \(a^{L-T}_{\hat{\bf{v}}} = 0\).
However, the effect of the even zonal harmonics on the acceleration
vector of a satellite is given by \(\bf{a} = {\bf{\nabla}} U\) and
the terms of the potential \(U\) generated by the mass, \(M\), of
the central body and by the even zonal harmonics, $J_{2n}$, are
\cite{kaula}: \(U = {M \over r} \, [ 1 - J_2 \, ({R_\oplus \over
r})^2 P_{20}(\sin \phi) - J_4 ({R_\oplus \over r})^4 P_{40}(\sin
\phi) - ...]\), where \(J_{2}\) and \(J_{4}\) are the
non--normalized even zonal harmonic coefficients of degree 2
(quadrupole) and 4, $R_\oplus$ is the Earth's equatorial radius and
\(P_{20}(\sin \phi)\) and \(P_{40}(\sin \phi)\) are the Legendre
associated functions. For a polar satellite with a circular orbit we
then have \(a^{J_{2n}}_\bot = 0\), \(a^{J_{2n}}_{\hat{\bf{r}}} \neq
0\) and \(a^{J_{2n}}_{\hat{\bf{v}}} \neq 0\). Then, for a polar
satellite with a circular orbit the satellite acceleration generated
by the even zonal harmonics is just orthogonal to satellite
acceleration generated by the gravitomagnetic field (i.e. the
Lense--Thirring effect). In conclusion, the static even zonal
harmonics determined by a polar satellite with a circular orbit are
independent on the a priori value used for the Lense--Thirring
effect.

Indeed, the orbit of the GRACE satellites is nearly circular with
eccentricity $< 0.005$ (the same is true for the CHAMP satellite
with eccentricity = 0.004) and basically polar with inclination =
89$^{\circ}$ (the same is true for the CHAMP satellite with
inclination = 87.3$^{\circ}$), thus the values of the even zonal
harmonics determined by the GRACE orbital perturbations are
substantially independent on the a priori value of the
Lense--Thirring effect. This  also applies to the CHAMP polar
satellite and to the previous CHAMP and GRACE only models. The small
deviation from a polar orbit of the GRACE satellite, that is \(1.7
\times 10^{-2}\) rad, gives only rise, {\it at most}, to a very
small correlation with a factor \(1.7 \times 10^{-2}\). For CHAMP
this correlation is, {\it at most}, $\sim$ \(4.7 \times 10^{-2}\).
In addition, since our results are independent of the error in the
quadrupole coefficient \(J_{2}\), any bias in \(J_{2}\) due to a
Lense--Thirring ``imprint'' is canceled in our combination (4).
However, the Lense--Thirring effect depends on the third power of
the inverse of the distance from the central body, i.e.,
\((1/r)^{3}\), and the \(J_{2}\), \(J_{4}\), \(J_{6}\) ... effects
depend on the powers \((1/r)^{3.5}\), \((1/r)^{5.5}\),
\((1/r)^{7.5}\) ... of the distance; then, since the ratio of the
semimajor axes of the GRACE satellites to the LAGEOS' satellites is
\(\sim \; {6780 \over 12270} \cong {1 \over 1.8}\), any conceivable
``Lense--Thirring imprint'' on the spherical harmonics at the GRACE
altitude becomes quickly, with increasing distance, a negligible
effect, especially for higher harmonics of degree \(l \geq 4\).
Therefore, any conceivable ``Lense--Thirring imprint'' is negligible
at the LAGEOS' satellites altitude. The same applies to the CHAMP
polar satellite. In addition, in \cite{ciucqg97}, it was proved with
several simulations that by far the largest part of this ``imprint''
effect is absorbed in the by far largest coefficient \(J_{2}\).

In conclusion, any error due to a conceivable ``imprint'' of the
Lense--Thirring effect on the Earth's even zonal harmonics of the
EIGEN-GRACE02S model (and of the other CHAMP and GRACE only models)
is negligible on the combination (4) of the nodes of LAGEOS and
LAGEOS II and therefore in our measurement of the Lense-Thirring
effect.

\section{A brief history of proposed experiments and measurements of
the Lense-Thirring effect using the nodes of the LAGEOS satellites}

The measurement reported in Nature \cite{ciupav} uses the {\it
nodes} of the two laser ranged satellites LAGEOS and LAGEOS 2 in
order to cancel the effect of the first even zonal harmonic
coefficient, $J_2$, of Earth and to measure the Lense-Thirring
effect. Furthermore, it uses the accurate model EIGEN-GRACE02S of
the Earth gravity field developed by GFZ of Potsdam with the data of
the NASA GRACE satellites.

The key idea of  using the nodes of two laser ranged satellites
of LAGEOS type to measure the Lense-Thirring effect was published
for the first time in 1984-86 \cite{ciu84,ciu86}).

The idea to use the nodes of N satellites of LAGEOS type to cancel
the effect of the first N-1 Earth even zonal harmonics and to
measure the Lense-Thirring effect was published for the first time
in 1989 \cite{ciu89}. The measurement of the Nature paper is simply
the case of  N=2. Indeed, in this 1989 paper on page 3102, fourth
line, in order to measure the Lense-Thirring effect and to cancel
the even zonal harmonics uncertainties, it is proposed (see also
\cite{ciuwhe}, on page 336): ``For $J_2$, this corresponds from
formula (3.2), to an uncertainty in the nodal precession of 450
milliarcsec/year, and similarly for higher $J_{2n}$ coefficients.
Therefore the uncertainty in $\dot \Omega_{Class}^{Lageos}$ is more
than ten times larger than the Lense-Thirring precession. A solution
would be to orbit several high-altitude, laser-ranged satellites,
similar to LAGEOS, to measure $J_2, J_4, J_6$ etc, and one satellite
to measure $\dot \Omega_{Lense-Thirring}$ ''. Thus, the case of the
Nature paper is just the one with two satellites. At that time the
error due to the even zonal harmonics was quite larger due to the
much less accurate Earth gravity models available at that time and
the LAGEOS 2 satellite was not yet launched (it was launched in
1992).

 {\it In particular, the idea, to use
the two nodes of the satellites LAGEOS  and LAGEOS 2 to measure the
Lense-Thirring effect}, together with the corresponding formula also
using the perigee of LAGEOS 2, was published for the first time in
1996 \cite{ciu96}. See, e.g., the formula (15) on page 1717 of ref.
\cite{ciu96}: ``$\delta \dot{\Omega}_I + k_1 \delta
\dot{\Omega}_{II} + k_2 \delta \dot{\omega}_{II} = \mu (31 + 31.5
k_1 - 57 k_2) \, milliarcsec/year  \, +  \, [other \, error \,
sources \, (\delta C_{60}, \delta C_{80}, ..., \delta I_I, \delta
I_{II})],$'' in the slightly different notation of the Nature paper
this is exactly eq. (1) of \cite{ciupav} where 31 and 31.5 are
written in the Nature paper: $\dot \Omega_I^{Lense-Thirring}$ and
$\dot \Omega_{II}^{Lense-Thirring}$ and $\delta C_{60}, \delta
C_{80}$, ... are just written $\Sigma \delta J_{2n}$, and $k_2 = 0$,
i.e. there is no use of the perigee, but just of the nodes, {\it
according to what we suggested in the above 1989 paper \cite{ciu89}}
and similar following papers. {\it All} the relevant LAGEOS and
LAGEOS 2 nodal rate coefficients of the $J_{2n}$, up to $J_4$, were
explicitly given in this 1996 paper.

In \cite{pet} various combinations of the orbital elements of the
satellites LAGEOS, LAGEOS 2 and LAGEOS 3 were also studied.

Finally, the combination (1) of ref. \cite{ciupav}, using the nodes
only, with the explicit value of the $k_1$ coefficient (that is
however an {\it absolutely trivial step} using the above formula and
the coefficients given in \cite{ciu96}, since it corresponds to set
$k_2 = 0$, according to what explained in \cite{ciu89} and
\cite{ciuwhe}, and to carry out the following highly trivial
arithmetics: $k_1 \cong - (-9.3/17.2) \cong 0.54$, where the
coefficients $K_{20}^I \cong -9.3 \cdot 10^{11}$ milliarcsec/yr and
$K_{20}^{II} \cong 17.2 \cdot 10^{11}$ milliarcsec/yr were
explicitly given in \cite{ciu96}) was  presented and published in
the SIGRAV 2002 Villa Mondragone (Rome) School proceedings
\cite{ciu02}, presented at a plenary talk at the Marcel Grossmann
meeeting in Rio de Janeiro in July 2003 (to appear in the
proceedings \cite{ciu03}) and at a number of other meetings since
2002. In almost every talk given by one of us from 1996 to 2004, it
was discussed the possibility of using the nodes only of LAGEOS and
LAGEOS 2 to just cancel the $J_2$ effect and measure the
Lense-Thirring effect. Therefore, the author of \cite{Iorio} seems
to have just rediscovered in 2003-2004 what already published by one
of us many years ago.

 The use of the GRACE-derived gravitational models, when available, to
measure the Lense-Thirring effect with accuracy of a few percent
was, since many years, a well known possibility to all the
researchers in this field and was presented by one of us during the
SIGRAV 2000 conference \cite{ecp} and published in its proceedings,
and was published by Ries et al. in the proceedings of the 1998
William Fairbank conference and of the 2003 13th Int. Laser Ranging
Workshop \cite{rie,rieWF}, as very clearly reported in the
Supplementary Discussion of \cite{ciupav} where it is written: ``...
(see also ref. [10] by Ries et al.-2003 concluding that, in the
measurement of the Lense-Thirring effect using the GRACE gravity
models and the LAGEOS and LAGEOS 2 satellites: ``a more current
error assessment is probably at the few percent level'') ...''.

In conclusion, all the claims of the author \cite{Iorio} are simply
lacking of any rational basis: above is shown how much work was
already published on this topic before the author of \cite{Iorio}
even began to produce any of his paper on this topic and to
rediscover some earlier results (the papers reported above are just
a small subset of all the work published by us in the last 20
years). To avoid the misunderstandings of \cite{Iorio}, it would
have just been a matter of very carefully reading the previously
existing literature on this subject!


Let us finally briefly comment on some historical milestones
in regard to the error analysis of
the Nature paper and to the uncertainties in the perturbations
of  the nodes of the LAGEOS satellites.

The error analysis of our 2004-measurement of the Lense-Thirring
effect with the LAGEOS and LAGEOS 2 satellites is substantially the
same as the error analysis of the LAGEOS III/LARES experiment that
was carried out in a very large number of papers and studies of
NASA, ASI and ESA \cite{NASA-ASI}. In particular, in 1989, this
analysis was performed under the supervision of an official NASA
committee chaired by Peter Bender, with members J. Anderson, J.
Armstrong, J. Breakwell, D. Christodoulidis, F. Everitt, E. Guinan,
R. Hellings, R. King, I. Shapiro, D. Smith and R. Weiss; then, in
1994, J. Ries et al., published a revised error budget of the LAGEOS
III experiment, with a total error of 4 $\%$ [this 1994 error
analysis of Ries et al. is reported with its reference in the
Supplementary Discussion of \cite{ciupav}] and, in 1998, with a
total error of 3$\%$-4$\%$ [see: \cite{ries2} in \cite{ciup}], as
reported in the Supplementary Discussion of \cite{ciupav}.

For differences and similarities between the error analysis of the
proposed LAGEOS III/LARES experiment and the one of the recent
measurement of the Lense-Thirring effect with LAGEOS and LAGEOS 2
\cite{ciupav}, see the Supplementary Discussion of \cite{ciupav} and
ref. \cite{ciupavper}. Here we just stress that {\it the studies of
the LAGEOS III/LARES experiment were fundamental milestones in this
field, indeed they analyzed in details all the main modelling errors
in the nodal rates of LAGEOS type satellites.}

\section{On the use of the mean anomaly
and on the use of Jason to measure the Lense-Thirring effect
proposed in (Iorio 2004)}

Besides the deep misjudgements and miscalculations of the impact of
$\dot J_4$ and of any ``Lense-Thirring imprint'' in the uncertainty
of our measurement of frame-dragging, one of the most profound
mistakes and misunderstandings of \cite{Iorio} is the proposed use
of the mean anomaly of a satellite to measure the Lense-Thirring
effect (in some previous paper by the same author the use of the
mean anomaly was also explicitly proposed; see e.g. L. Iorio "Some
comments on the recent results about the measurement of the
Lense-Thirring effect in the gravitational field of the Earth with
the LAGEOS and LAGEOS II satellites" arXiv:gr-qc/0411084 $\bf{v5 \;
19 \; Apr \; 2005}$, however its latest versions, after the online
publishing of our present paper in New Astronomy on the 2nd of May
2005, have been amended of this mistake). In \cite{Iorio}
(arXiv:gr-qc/0411024 $\bf{v9 \; 19 \; Apr \; 2005}$, however its
latest versions, after the online publishing of our present paper in
New Astronomy on the 2nd of May 2005, have been amended of this
mistake) is written:

``The problem of reducing the impact of the mismodelling in the even
zonal harmonics of the geopotential with the currently existing
satellites can be coped in the following way. Let us suppose we have
at our disposal N $(N > 1)$ time series of the residuals of those
Keplerian orbital elements which are affected by the geopotential
with secular precessions, i.e. the node, the perigee and the mean
anomaly: let them be $\Psi^A$, A = LAGEOS, LAGEOS II, etc. Let us
write explicitly down the expressions of the observed residuals of
the rates of those elements $\delta \dot{\Psi}^A_{obs}$ in terms of
the Lense–Thirring effect $\dot{\Psi}^A_{LT}$, of N - 1 mismodelled
classical secular precessions $\dot{\Psi}^A_{l} \delta J_l$ induced
by those even zonal harmonics whose impact on the measurement of the
gravitomagnetic effect is to be reduced and of the remaining
mismodelled phenomena $\Delta$ which affect the chosen orbital
element:

\begin{displaymath}
\delta \dot{\Psi}^A_{obs} = \delta \dot{\Psi}^A_{LT} + \Sigma
\dot{\Psi}^A_{l} \delta J_l + \Delta^A,
\end{displaymath}

\begin{displaymath}
 A = LAGEOS, LAGEOS II, ...   \;\;\;\;\;   (2)
\end{displaymath}

The parameter$^7 \; \mu$ is equal to 1 in the General Theory of
Relativity and 0 in Newtonian mechanics. The coefficients
$\dot{\Psi^A_{l}}$ are defined as

\begin{displaymath}
 \dot{\Psi}_{l} = {{\partial \dot{\Psi}_{class}} \over {\partial J_l}} \;\;\;\;\;   (3)
\end{displaymath}

and have been explicitly worked out for the node and the perigee up
to degree $l = 20$ in Iorio (2002b, 2003a); they depend on some
physical parameters of the central mass (GM and the mean equatorial
radius $R$) and on the satellite's semimajor axis $a$, the
eccentricity $e$ and the inclination $i$. We can think about Eq. (2)
as an algebraic non-homogeneuous linear system of N equations in N
unknowns which are $\mu$ and the N - 1 $\delta J_l$: solving it with
respect to $\mu$ allows to obtain a linear combination of orbital
residuals which is independent of the chosen N - 1 even zonal
harmonics.''

This is simply a nonsense paragraph: let us, for example, consider a
satellite at the LAGEOS altitude, the Lense-Thirring effect on its
mean longitude is of the order of 2 meters/yr, however
the mean longitude change is about $1.8 \cdot 10^{11}$ meters/yr.
Thus, from Kepler's law, the Lense-Thirring effect corresponds to a
change of the LAGEOS semi-major axis of about 0.09 millimeters.
Since, even a high altitude satellite such as LAGEOS showed a
semimajor axis change of the order of  1 millimeter/day, due to
atmospheric drag and to the Yarkoski-Rubincam effect (because of
atmospheric drag, the change of semimajor axis and mean motion is
obviously much larger for lower altitude satellites), and since the
present day $precision$ of satellite laser ranging is, even in the
case of the best SLR stations, of several millimeters, {\it it is a
clear nonsense to propose a test of the Lense-Thirring effect based
on using the mean anomaly of ANY satellite, mean anomaly largely
affected by non-conservative forces.}

In \cite{Iorio} and in one of Iorio's recent papers, it is proposed
to use the data collected from the satellite mission JASON, to
accurately measure the Lense-Thirring effect. Let us highlight some
of the major obstacles that one would face in trying to implement
such a proposal. A more detailed work discussing this proposal, the
use of the mean anomaly and other highly unfeasible proposals of the
author of \cite{Iorio} will be the subjects of a following paper.

First of all, Jason is a ``low Earth orbit'' satellite, at a 1330 km
altitude, and experiences by far larger perturbations than the
LAGEOS satellites from atmospheric drag, albedo and solar radiation
pressure and from all gravitational wavelengths. The complicated
shape of the satellite makes impossible to precisely
``forward-model'' all the non-conservative forces acting on the
satellite which generate perturbations that are making JASON a
totally unsuitable target for making a measurement of such a tiny
effect as the Lense-Thirring drag.

JASON is an oceanographic mission which measures radar ranges from
the satellite to the surface of the ocean with a precision of about
4 cm. To make use of these observations for oceanographic research,
the requirement on the orbital accuracy is stringent but certainly
not extraordinary. A couple of centimeters of radial error are
adequate, although recent results indicate that we may be closer to
1 cm \cite{luth}. The JASON bus is a very large, asymmetrical one
and it was not designed to support millimeter ranging work, since
oceanography is still trying to rationalize much larger signals. In
addition to the large size and poorly defined shape in terms of
properties, etc., the satellite undergoes nearly continuous  orbital
adjustments to maintain its ground-track  within $\pm$ 1 km off the
accepted mean ground-track of the Topex/Poseidon mission and in this
process its ideal orbit as a test particle in the Earth
gravitational field changes by thrusting and maneuvering; this is
never easily controlled and it cannot be taken into account in
computing a perfectly reconstituted orbit. The shape of the
satellite bus and the continuous motion of parts on the bus (e.g.,
solar panels) as well as the satellite itself (yawing or not, to
maintain orientation with respect to the Sun) generate further
accelerations that are unknown and extremely difficult to account
for, except by solving for what we call ``ad hoc'', non-physically
meaningful, accelerations at certain intervals and that, however,
will absorb signals such as the L-T effect. To avoid these problems
a large international group of researchers proposed the
LARES/WEBERSAT mission \cite{ciup} and, in doing so, designed a new
structure for the LARES satellite, a very clean cannonball
satellite, similar to LAGEOS but improved for what regards the
effect of the non-gravitational perturbations. In any case, one
wonders why the author of \cite{Iorio} has not proposed the use of which not only have by far better
shape and orbital stability compared to JASON but they also carry
ultra precise accelerometers that measure all non-conservative
accelerations, so that they are in practice ``free-falling''
particles in vacuum, at least to the extent that is covered by the
accuracy of these instruments.

In conclusion, not only all the criticisms in Iorio's paper are
completely unfounded and misdirected and our total error bubget is
fully confirmed to be between 5 $\%$ and 10 $\%$ of the
Lense-Thirring effect, but Iorio's paper shows a profound lack of
understanding and knowledge of the real, practical, data analysis of
laser ranging observations and of the actual modeling of the Earth
gravity field.

\newpage

\noindent {\Large \bf Acknowledgements} \\

\noindent One of us (I.C.) is very grateful to Richard Matzner for
reading the manuscript and for helpful suggestions.

\end{document}